\documentstyle [twocolumn,seceq] {jpsj}
\input epsf.sty
\def\runtitle{Spin dynamics simulations of excitations...}
\def\runauthor{{\sc Landau}, {\sc Tsai}, and {\sc Bunker}}
\newcommand{\beq}{\begin{equation}}
\newcommand{\eeq}{\end{equation}}
\newcommand{\beqs}{\begin{eqnarray}}
\newcommand{\eeqs}{\end{eqnarray}}

\title{\bf Spin dynamics simulations of excitations and critical
dynamics in RbMnF$_3$ }

\author{D. P. {\sc Landau}, Shan-Ho {\sc Tsai}
 and Alex {\sc Bunker} \footnote{Present address: Max Planck Institute
 for Polymer Research, Ackermann Weg 10, Mainz, Germany D-55021-3148}
}

\inst{
Center for Simulational Physics \\
The University of Georgia \\
Athens, GA 30602
}

\recdate
{
\today
}
\abst
{
Spin-dynamics simulations have been used to investigate the dynamic behavior 
of RbMnF$_3$, treating it as a classical Heisenberg antiferromagnet on a
simple cubic lattice. Time-evolutions 
of spin configurations were determined numerically from coupled equations of motion 
for individual spins using a new algorithm  
which is based on Suzuki-Trotter decompositions of exponential operators. 
The dynamic structure factor was calculated from the space- and
time-displaced spin-spin correlation function. The crossover from hydrodynamic 
to critical behavior of the dispersion curve and spin-wave half-width was 
studied as the temperature was increased towards the critical value. The
dynamic critical exponent was estimated to be $z=(1.43\pm 0.03)$, which is 
slightly lower than the dynamic scaling prediction, but in good agreement with 
a recent experimental value. Comparisons are made of both the 
dispersion curve and the lineshapes obtained from our simulations with very 
recent experimental results for RbMnF$_3$ are presented. 
}

\kword
{
spin dynamics, critical dynamics
}

\vspace{1cm}

\begin{document}
\sloppy
\maketitle

\section{Introduction}

RbMnF$_3$ has been the subject of numerous experimental and theoretical
investigations since it is a good physical realization of an isotropic
three-dimensional Heisenberg antiferromagnet. 
Early experimental studies \cite{pickart,teaney,kornblit,windsor} showed that 
the Mn$^{2+}$ ions, with spin $S=5/2$, form a simple cubic lattice structure 
with a nearest-neighbor exchange constant $J^{exp}=(0.58\pm 0.06)$ meV and a 
second-neighbor interaction constant of less than $0.04$ meV [both defined 
using the exchange constant to be shown in Eq. (\ref{hamq}), the normalization
used here differs from that of Ref.\citen{windsor} by a factor of two].
Magnetic ordering with antiferromagnetic alignment of spins occurs below
the critical temperature $T_c = 83K$. The magnetic anisotropy is very low, 
about $6\times 10^{-6}$ of the exchange field, and no deviation from cubic 
symmetry was seen at $T_c$ \cite{teaney62,teaney63}. 

Both the static properties 
and the dynamic response of RbMnF$_3$ have been examined through neutron
scattering experiments.  Windsor and co-workers\cite{windsor} looked at spin-
waves at low temperatures and mapped out the dispersion curve.
The early work of Tucciarone {\it et al} \cite{tucciarone} found that
in the critical region the neutron scattering function has a central peak 
(peak at zero frequency transfer) and a spin-wave peak. Later experiments 
by Cox {\it et al} \cite{cox} observed a small central peak below 
$T_c$ as well. The more recent study by 
Coldea {\it et al} \cite{coldea} also found central peaks for $T\le T_c$, in 
agreement with previous work. From the theoretical side,
renormalization-group (RNG) below $T_c$ \cite{mazenkoTb} predicts
spin-wave peaks, and a central peak in the longitudinal component of
the neutron scattering function; however, at $T_c$
both renormalization-group \cite{mazenkoTc} and mode-coupling \cite{cuccoli} 
theories predict only the presence of a spin-wave peak.  The experimentally
observed central peak is thought to be caused by spin diffusion resulting from 
nonlinearities in the dynamical equations \cite{tucciarone}.
Coldea {\it et al} \cite{coldea} also obtained the most precise experimental 
estimate of the dynamic critical exponent, $z=(1.43\pm 0.04)$.  This is 
slightly smaller than the predicted value \cite{hohenhal}
of $z=1.5$ for an isotropic Heisenberg antiferromagnet in $d=3$ dimensions.

Extensive Monte Carlo studies measured the static properties of 
classical Heisenberg magnets but could not examine the true dynamics of the
systems.  Several large-scale spin-dynamics simulations have probed the
behavior of various classical systems \cite{kun, alex}; however, there are 
no direct comparisons to experimental data for physical systems.
In the present work we report large-scale simulations of the dynamic
behavior of the Heisenberg antiferromagnet on a simple cubic lattice, and make 
direct comparison with experimental data. 

\section{Model and Methods}

The classical Heisenberg antiferromagnet is defined by the Hamiltonian
\beq
{\cal H} = J \sum_{<{\bf rr'}>} {\bf S_r}\cdot {\bf S_{r'}},
\label{ham}
\eeq

\noindent
where ${\bf S_r}=({S_{\bf r}}^x,{S_{\bf r}}^y,{S_{\bf r}}^z)$ is a three-dimensional
classical spin of unit length at site ${\bf r}$ and $J>0$ is the 
antiferromagnetic coupling constant between nearest-neighbor pairs of spins. 
All simulations were performed using $L\times L\times L$ simple cubic lattices 
with periodic boundary conditions. The dynamics of the spins are governed 
by the coupled equations of motion \cite{gerling}
\beq
\frac{d}{dt}{\bf S_r} = - {\bf S_r} \times J\sum_{<{\bf rr'}>} {\bf S_{r'}},
\label{eqofmotion}
\eeq

\noindent
and the time dependence of each spin can be determined from the
integration of these equations.

The dynamic structure factor $S({\bf q},\omega)$ for momentum transfer
${\bf q}$ and frequency transfer $\omega$ can be measured by
inelastic neutron scattering experiments and is given by
\beq
S^k({\bf q},\omega)=\sum_{\bf R} e^{i {\bf q}\cdot {\bf R}}
\int_{-\infty}^{+\infty} e^{i\omega t} C^k({\bf R},t) \frac{dt}{\sqrt{2\pi}},
\eeq

\noindent
where $\bf R = {\bf r} - {\bf r'}$,  $C^k({\bf R},t)$ is
the space-displaced, time-displaced correlation
function, with $k=x, y,$ or $z$, and
\beq
C^k({\bf R},t) =\langle {S_{{\bf r}}}^k(t){S_{{\bf r'}}}^k(0)\rangle-
\langle {S_{{\bf r}}}^k(t)\rangle\langle {S_{{\bf r'}}}^k(0)\rangle.
\eeq

\noindent
In the case of antiferromagnets, the wave-vectors are measured with respect to the 
$(\pi,\pi,\pi)$ point which corresponds to the Brillouin zone center.
Note that in the [1,1,1] and [1,0,0] directions the respective
first Brillouin zone boundary wave-vectors are $(\pm \pi/2,\pm \pi/2,\pm \pi/2)$ 
and $(\pm\pi,0,0)$.

Using Monte Carlo and spin-dynamics methods\cite{kun,alex,gerling}, 
we simulated the simple-cubic classical Heisenberg 
antiferromagnet with $12\le L\le 60$ at the critical temperature\cite{kunTc} 
$T_c=1.442929(77)J$ as well as below $T_c$. (We use units in which Boltzmann's
constant $k_B=1$.)  Equilibrium configurations were generated using a hybrid 
Monte Carlo method \cite{kun,alex} and the coupled equations of motion were 
then integrated numerically, using these states as initial spin configurations.  
Numerical integrations were performed to a maximum time $t_{max} \le 1000J^{-1}$, 
using a time step of $\Delta t$. The space-displaced, time-displaced
correlation function $C^k({\bf R},t)$ was computed for
time-displacements ranging from $0$ to $t_{cutoff}$ and extracted from an 
average over 40 to 80 different time starting points, evenly spaced by 
$10\Delta t$.  As many as $7000$ initial configurations were used, although 
for large lattices this was reduced to as few as $400$.  For $L=24$ at 
$T=0.9T_c$ the integration was carried out with a time step $\Delta t=0.01J^{-1}$
using a 4th-order predictor-corrector method.\cite{kun,alex}.  For other 
lattice sizes and temperatures, we used a new algorithm \cite{frank,krech} based 
on 4th-order Suzuki-Trotter decompositions of exponential operators, with a 
time step $\Delta t=0.2J^{-1}$ . The larger integration time step allowed us
to extend the maximum integration time to much larger values than was previously
possible. 

In order to reduce the computer resources needed we calculated partial 
spin sums ``on the fly''\cite{kun,alex}; however, data 
could then only be kept for the $(q,0,0)$, $(q,q,0)$ and $(q,q,q)$ directions
with $q$ determined by the periodic boundary conditions,
\beq
q=\frac{2\pi n}{L},\qquad n=\pm 1,\pm 2,...,\pm (L-1),L.
\label{qpbc}
\eeq

\noindent
Since all three Cartesian spatial directions are equivalent by symmetry, results
for $(q,0,0)$, $(0,q,0)$ and $(0,0,q)$ were averaged. Similarly, the same
operations carried out for the $(q,q,0)$ and $(q,q,q)$ directions were also 
averaged over the equivalent reciprocal lattice directions. 

For the Heisenberg ferromagnet the order parameter is the
total magnetization, which is a conserved quantity. Thus 
the dynamic structure factor $S({\bf q},\omega)$ can be separated
into a component along the axis of the
total magnetization (longitudinal component) and a transverse component; 
however, for the isotropic
antiferromagnet considered here the order parameter is not conserved and the
longitudinal and transverse components of $S({\bf q},\omega)$ cannot be 
separated in the simulation.  Henceforth we will use the term dynamic structure 
factor to refer to the average.

Two practical limitations on spin-dynamics techniques 
are the finite lattice size and the finite evolution time. The finite time cutoff can introduce
oscillations in $S({\bf q},\omega)$, which can be smoothed out by convoluting the spin-spin
correlation function with a resolution function in frequency. In 
neutron scattering experiments
the divergence of the neutron beam gives rise to an intrinsic Gaussian
resolution function in ${\bf q}$
and $\omega$ and the smoothed dynamic structure factor is

\beqs
\bar {S_{\xi}}^k({\bf q},\omega) &\equiv& \sum_{{\bf R}} e^{i{\bf q}\cdot
{\bf R}}\int_{-t_{cutoff}}^{+t_{cutoff}} e^{i\omega t} C^k({\bf R},t)
e^{-\frac{(t\delta_{\omega})^2}{2} } \frac{dt}{2\pi} \nonumber \\
&\approx & \frac{1}{\sqrt{2\pi}\delta_{\omega}}\int_{-\infty}^{+\infty}{S_{\xi}}^k({\bf q},\omega')
e^{ -\frac{(\omega -\omega')^2}{2\delta_{\omega}^2}} d\omega',
\label{gaussrf}
\eeqs

\noindent
where $\delta_{\omega}$ is a parameter characterizing the Gaussian resolution function and has
to be chosen properly so that effects due to the finite time cutoff can be neglected. 
The momentum dependent susceptibility, ${\bar{\chi}_{\xi}}^k({\bf q})$, is given by 
\beq
\int_{-\infty}^{\infty}\bar {S_{\xi}}^k({\bf q},\omega)\frac{d\omega}{2\pi}=
{\bar{\chi}_{\xi}}^k({\bf q}).
\eeq

Finite-size scaling theory \cite{kun,rapaport} can be used to extract the 
dynamic critical exponent $z$:  the divergence of the correlation length $\xi$ 
is limited by $L$ and the dynamic finite-size relations are given by
\beq
\frac{\omega\bar {S_L}^k({\bf q},\omega)}{{\bar{\chi}_L}^k({\bf q})}=
G(\omega L^z,qL,\delta_{\omega}L^z)
\eeq
and 
\beq
\bar\omega_m=L^{-z}\bar\Omega(qL,\delta_{\omega}L^z),
\label{omegam}
\eeq
where $\bar\omega_m$ is a characteristic frequency, defined as
\beq
\int_{-\bar\omega_m}^{\bar\omega_m}\bar {S_L}^k({\bf q},\omega)\frac{d\omega}{2\pi}
=\frac{1}{2}{\bar{\chi}_L}^k({\bf q}).
\eeq

For $t_{cutoff} \ge 400J^{-1}$ the oscillations 
in the dynamic structure factor were not
very significant.  Thus, we first estimate the dynamic critical exponent
$z$ without using a resolution function, i.e. we take $\delta_{\omega}=0$. In 
this case $z$ can be obtained from the slope of a graph of $\log(\omega_m)$ vs $\log(L)$ (where
$\omega_m$ is the characteristic frequency for $\delta_{\omega}=0$) 
if $qL$ is fixed and $L$ is large enough to be in the asymptotic-size regime. 

The effects of the small oscillations in $S(q,\omega)$ 
on the dynamic exponent $z$ can be evaluated by repeating the analysis using a
resolution function
so that the function $\bar\Omega(qL,\delta_{\omega}L^z)$ in Eq. (\ref{omegam}) 
is constant if $qL$ and $\delta_{\omega}L^z$ are fixed and
\beq
\bar\omega_m\sim L^{-z}.
\label{omegamLz}
\eeq
Because $\delta_{\omega}$ depends on $z$, this exponent had to be 
determined iteratively. An initial value $z^{(0)}$ was used to determine  
$\delta_{\omega}$ for different $L$, $\bar {S_L}^k({\bf q},\omega)$ and 
$\bar\omega_m$ were computed for different 
values of $L$ and $q$ with $qL$ held fixed (i.e. $n$ is constant) and a 
new estimate, $z^{(1)}$, was extracted from 
Eq. (\ref{omegamLz}). This procedure is repeated until the estimates
converge.

\section{Results\label{results}}
\subsection{Numerical data for $S({\bf q},\omega)$}

\begin{figure}[htb]
\vspace{-0.5cm}
%\vspace{1.5cm}
\centering
\leavevmode
\epsfxsize=3.0in
\begin{center}
\leavevmode
\epsffile{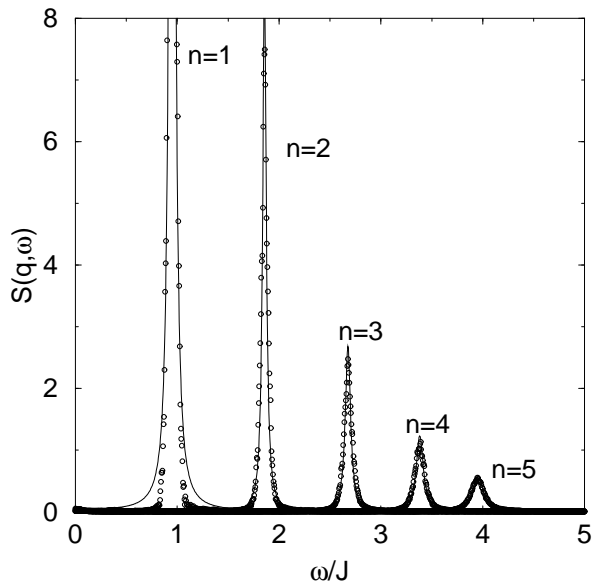}
\end{center}
\vspace{-1.8cm}
\caption{Dynamic structure factor $S({\bf q},\omega)$ from our simulations
for $L=20$ at $T=0.4T_c$, $q$ in the [100] direction. 
The symbols represent spin dynamics data and the solid line is a fit with the
Lorentzian function given in Eq. (\ref{lorentz}).}
\label{lowT}
\end{figure}

\begin{figure}
\vspace{-1.0cm}
\centering
\leavevmode
\epsfxsize=2.8in
\begin{center}
\leavevmode
\epsffile{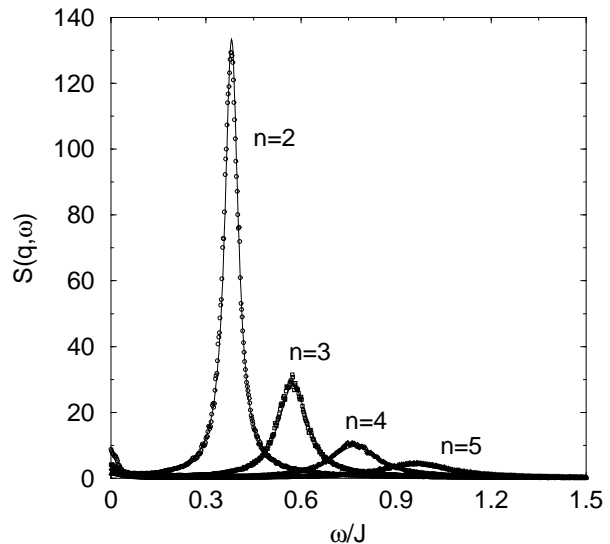}
\end{center}
\vspace{-2.0cm}
\caption{ Dynamic structure factor $S({\bf q},\omega)$ 
for $L=60$ at $T=0.9T_c$, and $q$ in the [100] direction. 
The symbols represent spin dynamics data and the solid line is a fit with the
Lorentzian function given in Eq. (\ref{lorentz}).}
\label{lshape09}
\end{figure}

\begin{figure}
\vspace{-2.0cm}
\centering
\leavevmode
\epsfxsize=2.7in
\begin{center}
\leavevmode
\epsffile{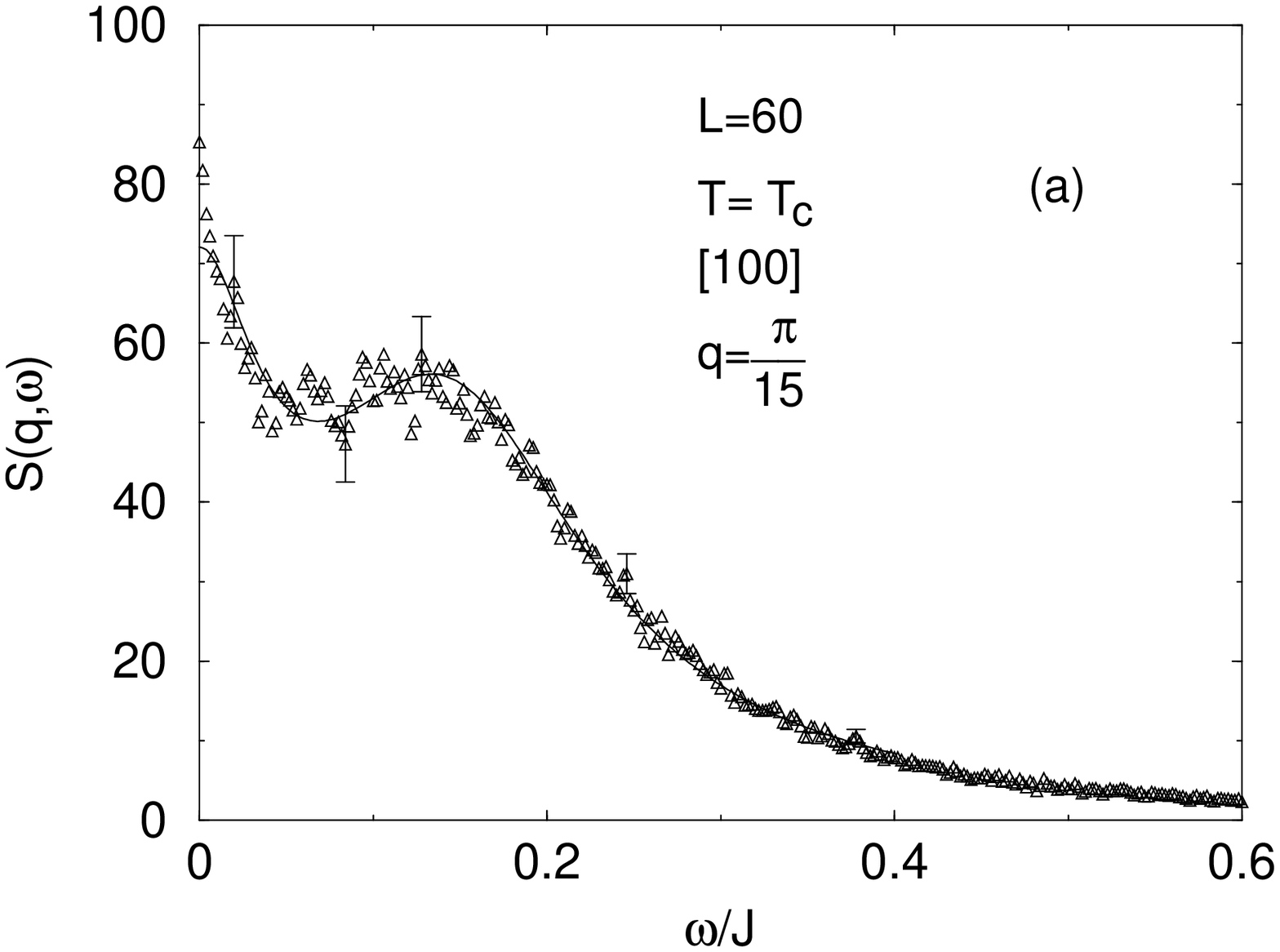}
\end{center}
\vspace{-3.5cm}
\begin{center}
\leavevmode
\epsfxsize=2.7in
\epsffile{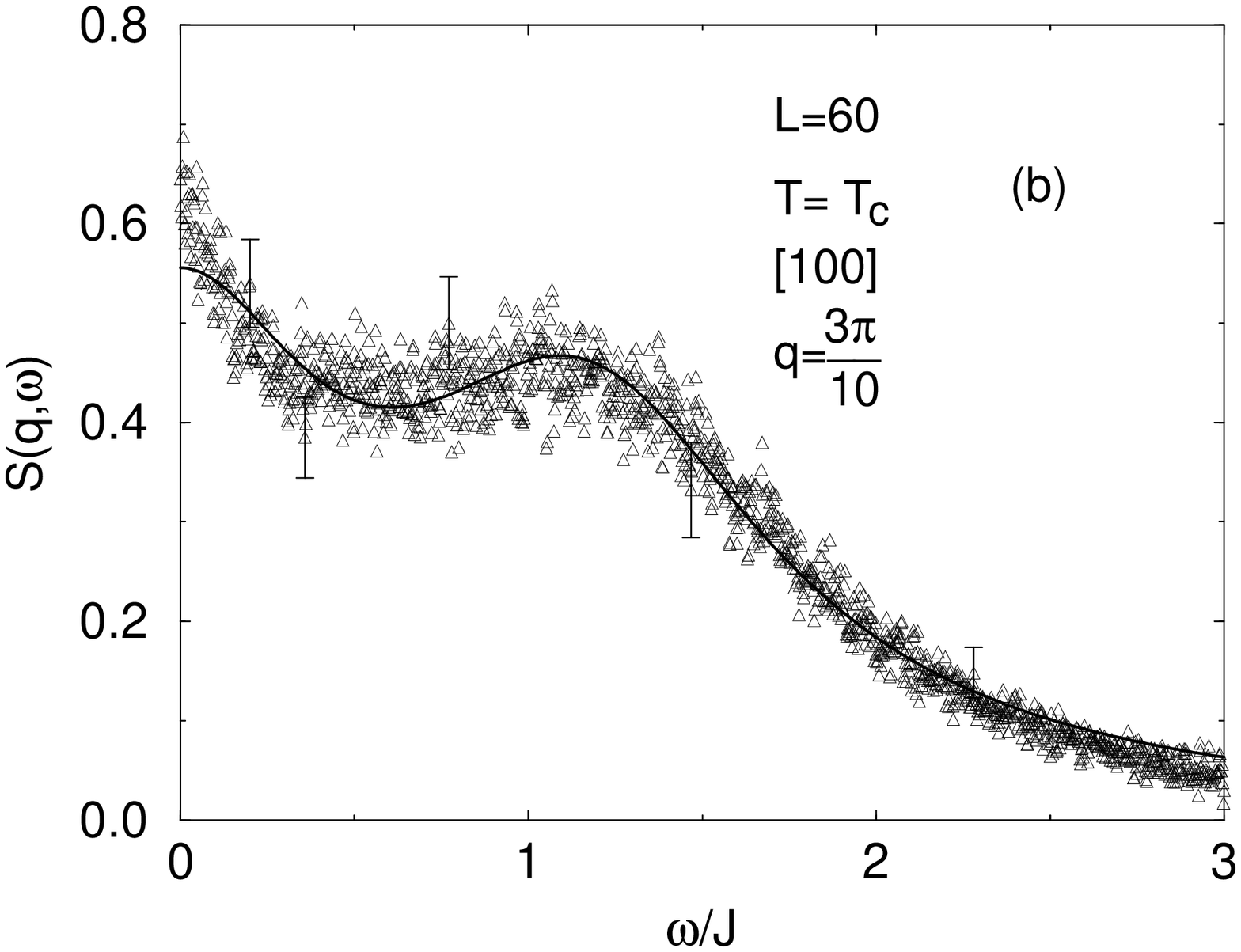}
\end{center}
\vspace{-1.3cm}
\caption{Dynamic structure factor $S({\bf q},\omega)$ at $T=T_c$:  (a) $q=\pi/15$ 
and (b) $3\pi/10$ in the [100] direction.  The symbols represent spin dynamics 
data for $L=60$ and the solid line is a fit to the sum of Lorentzians given 
in Eq. (\ref{lorentz}). Error bars are only shown for a few typical points;
at high frequency they are of the size of the fluctuations in the data.}
\label{lshapeTc}
\end{figure}

For $T\le T_c$ our results for the dynamic structure factor  
show a spin-wave and a central peak.  At
low temperatures the central peak is barely visible and very narrow 
spin wave peaks are the dominant feature, see Fig. \ref{lowT}
(finite-size effects are evident for $n=1$).
In Fig. \ref{lshape09} we show lineshapes for lattice size $L=60$ at 
$T=0.9T_c$ and several q-values in the [100]  
direction. As $q$ increases, the central peak broadens and its
relative amplitude increases.  As the temperature is increased, the central peak
grows, and at $T_c$ the central peak is even stronger than the spin wave peak.
Fig. \ref{lshapeTc} shows lineshapes for $L=60$ and $q=\pi/15$ and $3\pi/10$ 
in the [100] direction.  Clearly oscillations due to the finite $t_{cutoff}$
are negligible; therefore, in our lineshape analysis we did not use
a resolution function. For direct comparison with experimental
%\noindent
data we convoluted our results with a Gaussian resolution function 
with the same width as in the experiment.  The structure 
in the lineshapes discussed here is much larger than our resolution in
frequency.
\noindent
Below $T_c$, previous theoretical \cite{mazenkoTb} and experimental \cite{coldea} 
studies provided the comparison of the position and the half-width 
of the spin-wave 
and central peaks by fitting the lineshape to a Lorentzian form
\beq
S({\bf q},\omega)=\frac{A\Gamma_1^2}{\Gamma_1^2+\omega^2}+\frac{B\Gamma_2^2}
{\Gamma_2^2+(\omega+
\omega_s)^2}+\frac{B\Gamma_2^2}{\Gamma_2^2+(\omega-\omega_s)^2}
\label{lorentz}
\eeq
where the first term corresponds to the central peak and the last two terms are
from the spin-wave creation and annihilation peaks at $\omega=\pm\omega_s$. 

%\vspace{1.0cm}
\noindent
For $T=0.9T_c$ Lorentzian lineshapes fit our results well for small values of 
$q$, except for the smallest value, $q=2\pi/L$, in the [100] direction for 
which finite size effects are apparent.  For large
values of $q$ the Lorentzian form given in Eq. (\ref{lorentz}) does not
fit the data, especially at high frequency. In general, the fitted 
parameters varied when different frequency ranges were used in the fit by an 
amount which was often larger than the statistical error in the fitted 
parameters obtained from the fit using a single frequency range.
Therefore, for $T=0.9T_c$ we estimated the error in the fitted parameters 
by fitting the lineshapes using three different ranges of frequency and taking
the average. 
At $T_c$, renormalization-group theory (RNG) \cite{mazenkoTc} predicts a non-Lorentzian functional 
form for the spin-wave lineshape, which has been used along with a Lorentzian central peak
to analyze experimental data \cite{coldea}.
Since it is more complicated to perform fits to this RNG functional form and since the 
spin-wave peaks obtained from the simulations are more pronounced than in the experiment,
and thus less dependent on the fitted functional, we have fitted the lineshapes at $T_c$ to 
Lorentzians, as given in Eq. (\ref{lorentz}). Although 
obtaining a good fit to our data at $T_c$  was more difficult than below $T_c$, the 
resulting fits are still reasonable. Unlike for $T=0.9T_c$, at $T_c$ the lineshape 
parameters used in the analysis below are the values obtained from the fit 
to only one frequency range, which was the one that gave the best fit. 
The actual error in the fitted parameters at $T_c$ should be larger (by up
to a factor of 5) than the error bars shown in the figures below. 
Illustrations of the fits using Eq. (\ref{lorentz}) to the simulated lineshapes 
at $T=0.9T_c$  and at $T_c$ for several values of $q$ are shown 
in Figs. \ref{lshape09} and \ref{lshapeTc}.

We also found very weak peaks in the high frequency tail of the spin-wave peaks. 
Using the spin-wave 
frequencies in the [100], [110] and [111] directions we could check that the 
position of these extra peaks corresponded to frequencies of two spin-wave 
addition peaks.  These extra structures in the lineshapes were particularly 
visible for the smallest $q$-values. 

\begin{figure}
\vspace{-1.7cm}
\centering
\leavevmode
\epsfxsize=2.8in
\epsffile{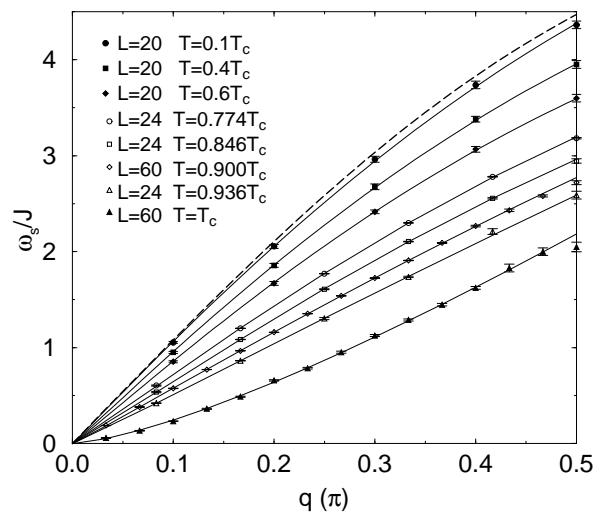}
\vspace{-1.3cm}
\caption{ Spin-wave dispersion relations for $T\leq T_c$,
in the [100] direction. The symbols represent spin-wave positions extracted from Lorentzian
fits to the lineshapes from the simulations, and the solid curves are fits of the dispersion
relations at different temperatures to Eq. (\ref{wsfita0q2}).}
\label{wqfig}
\end{figure}

\begin{figure}
\vspace{-2.0cm}
\centering
\leavevmode
\epsfxsize=3.0in
\epsffile{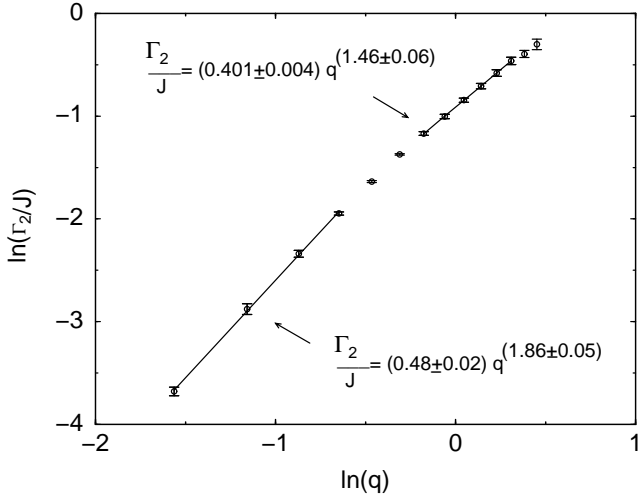}
\vspace{-1.3cm}
\caption{Log-log graph of the half-width of the spin-wave peak extracted from 
Lorentzian fits to the lineshapes obtained from simulations for $L=60$ and 
$T=0.9T_c$ in the [100] direction as a function of $q$.}
\label{hwq0p9}
\end{figure}

Fig.  \ref{wqfig} shows how the dispersion curve varies as the temperature 
increases from $T=0$ to $T_c$. Although the Lorentzian in Eq. (\ref{lorentz}) did not yield good 
fits to the lineshapes for larger 
values of $q$, the spin-wave peak positions could still be 
determined relatively accurately and the dispersion curve could be measured
up to $q=\pi/2$.  Well below $T_c$, the dispersion relation is linear for
small $q$, but as the temperature increases towards $T_c$, the dispersion 
relation changes gradually to power-law 
\beq
\omega_s=A_s q^x.
\label{wsfita0}
\eeq
For $T=0.9T_c$ and $L=60$ a fit to the smallest five $q$ values of the 
dispersion curve to Eq. (\ref{wsfita0}) yielded $x= 1.017\pm 0.003$. If larger 
values of $q$ are included in the fit, the exponent decreased slightly. 
The sensitivity of the fitted exponent to the particular form of the 
fitted function was examined by including a quadratic term, i.e., 
\beq
\omega_s=A_s q^x + B_s q^2,
\label{wsfita0q2}
\eeq
yielding a similar value $x=1.020\pm 0.003$. When larger values of $q$ were 
included in the fits, Eq. (\ref{wsfita0q2}) tended to yield smaller 
$\chi^2$'s per degree of freedom than Eq. (\ref{wsfita0}).  The dispersion 
curve for $T=T_c$ and $L=60$ fitted to Eq. (\ref{wsfita0}) yielded an exponent 
of $x=1.38\pm 0.01$ when the smallest 12 values of $q$ were included in the fit. As the larger
$q$ were excluded from the fit, the exponent increased slightly, tending 
towards $x\simeq 1.40$.  When only the smallest few values of $q$ were 
included in the fit, the exponent decreased again, reflecting the fact that 
we probed correlations between spins separated by larger distances, i.e. 
smaller $q$.  This reflected the finite size of the lattice (and thus of the 
correlation length), showing that the system is not at 
criticality. Hence, the exponent $x$ decreases towards unity. In contrast, 
large values of $q$ correspond to short distance (in the direct 
lattice space) spin-spin correlations, and the correlation length is much 
larger than the distance probed.  Our results at $T_c$ agree with recent 
experiment  \cite{coldea} which found $x=1.43\pm 0.04$ when 
the dispersion curve at $T_c$ was fitted to a power-law
relation of the form given in Eq. (\ref{wsfita0}).  The solid lines 
in Fig. \ref{wqfig} are fits to Eq. (\ref{wsfita0q2}); in general, these fits
gave lower values of $\chi^2$ per degree of freedom than fits to Eq. (\ref{wsfita0}).
In the critical region, dynamic scaling theory predicts \cite{dynscalhh} that 
the half-width of spin-wave
peaks behaves as $\Gamma_2 \sim q^{1.5}$, whereas for the
hydrodynamic regime the prediction from hydrodynamic theory\cite{hydrohh}  
is $\Gamma_2 \sim q^{2}$. 
%
%\vspace{1.0cm}
\noindent
The half-width of the spin-wave peaks at $T=0.9T_c$ and $L=60$ from our 
simulations is shown in Fig. \ref{hwq0p9}. We observed a crossover from  
$\Gamma_2 =(0.401\pm 0.004) q^{1.46\pm 0.06}$ for larger values of 
$q$ to the behavior $\Gamma_2 =(0.48\pm 0.02) q^{1.86\pm 0.05}$ for small 
values of $q$.  The behavior for relatively large $q$ 
agrees with dynamic scaling theory and with recent experiment 
\cite{coldea}. The exponent we obtained by fitting only small values of $q\;$ 
%
%\vspace{1.0cm}
is close to the hydrodynamic prediction. Thus, the spin-wave half-width 
reflected crossover between two different 
regimes, the critical and the hydrodynamic regions. This crossover 
is similar to the one observed in the 
dispersion curve at $T_c$, discussed above. For $T=T_c$ and
$L=60$ the spin-wave half-width also had a
power-law behavior which varied from approximately $q^{1.2}$ when the
12 smallest values of $q$ were included to $\sim q^{1.4}$ when only 
the smallest five wave-vectors were considered. In their recent 
experiment, Coldea {\it et al} \cite{coldea} found 
$\Gamma_2 = D q^{1.41\pm 0.05}$ for $0.77T_c\leq T < T_c$, 
and the coefficient $D$ increased with increasing temperature. 

As in the experiments, the dynamic structure factors from our simulations had
central peaks ($\omega = 0$) for $T\leq T_c$.  In contrast, 
RNG theory predicts a central peak in the longitudinal component 
of the dynamic structure factor only below $T_c$  \cite{mazenkoTb}, and 
none of the theories predict a central peak at $T_c$ \cite{mazenkoTc,cuccoli}. 
For $T=0.9T_c$ and $L=60$ fits of the central peak half-width to a simple
power law were poor, but a 
much improved fit was obtained by using the function $\Gamma_1= A_1+B_1q^{C_1}$, which allows
for a non-zero central peak width when $q$ vanishes. In these
fits the data for the smallest possible $q$, i.e. $n=qL/2\pi=1$, were 
excluded because of large finite-size effects. The fit including data for 
$q$ corresponding to $n=2$ until $n=7$ yielded 
$A_1\simeq 0.013\pm 0.001$, $B_1\simeq 0.120\pm 0.005$ and $C_1\simeq 2.4\pm 0.2$. As we
systematically included larger values of $q$ in the fits, these parameters decreased slightly.
At $T_c$ we also fitted the central peaks to Lorentzians, according to 
Eq. (\ref{lorentz}); however, these tended to yield curves with smaller 
amplitudes than the data. Since there 
is no theoretical prediction for the central peak, we have also tried to fit 
them with a Gaussian form but the result was much worse than with Lorentzians.

The lattice sizes that we used were all multiples of $12$ so there were
certain $q$-values which were common to all $L$. 
Lineshapes and spin-wave peak positions could be compared for different $L$ at a 
fixed value of $q$. At $T=0.9T_c$ we saw no significant finite-size effect
for $L\geq 24$; however, when we superimposed lineshapes at $T_c$ for  
fixed $q$, and different values of $L$, finite-size effects were noticeable for 
$L=24$. For larger $L$ the lineshapes were the same within the error bars.  

The dynamic critical exponent $z$ was extracted from finite-size scaling 
of $\bar\omega_m$. From an analysis without resolution function, or
equivalently $\delta_{\omega}=0$, and $n=1,2$, we estimated $L=30$ to be 
approximately the onset of the asymptotic-size regime and
$z=1.45\pm 0.01$ for $n=1$ and $z=1.42\pm 0.01$ for
$n=2$.  We also estimated the value of $z$ using a resolution function.
Several initial values of $z^{(0)}$ were used, and in all cases the exponent $z$ 
converged rapidly.
Our final estimate for the dynamic critical exponent is  $z=1.43\pm 0.03$.
Analyses of the characteristic frequency  $\bar\omega_m$ as a function of 
$L$ with and without a resolution function agreed closely.

\subsection{Comparison with experiment for RbMnF$_3$}

We now compare our results with the recent neutron scattering 
data of Coldea {\it et al} \cite{coldea}.  The Mn$^{2+}$ ions in
RbMnF$_3$ have spin $S = 5/2$ and interact via a quantum
Heisenberg Hamiltonian of the form
\beq
{\cal H} = J^{exp} \sum_{<{\bf rr'}>} {\bf S_r}^Q\cdot {\bf S_{r'}}^Q,
\label{hamq}
\eeq
where ${\bf S_r}^Q$ are spin operators with magnitude $|{\bf S_r}^Q|^2=S(S+1)$ 
and the interaction strength between pairs of nearest-neighbors was 
determined experimentally \cite{windsor} to be $J^{exp}=(0.58\pm 0.06)$ meV. 
In contrast, our simulations were performed on a classical  
Hamiltonian; however, quantum Heisenberg systems 
with large spin values ($S \geq 2$) have been shown to behave as classical 
Heisenberg systems where the spins are vectors of magnitude 
$\sqrt{S(S+1)}$ with the same interaction strength between pairs of nearest 
neighbors as in the quantum system \cite{classqu}.  Since our classical spins
were vectors of unit length, a normalization of the interaction strength 
$J$ from our simulation is needed, i.e.
\beq
J=J^{exp}S(S+1).
\label{jjexp}
\eeq 
Although this choice leaves the Hamiltonian unchanged, it modifies the 
equations of motions given above as Eq. (\ref{eqofmotion}). The dynamics of 
the classical system so defined is the same as the quantum system defined 
by the Hamiltonian in Eq. (\ref{hamq}) if one rescales the time, or
equivalently, the frequency. We obtain
\beq
\omega^{exp}=J^{exp}\sqrt{S(S+1)}\; \frac{w}{J},
\label{wwexp}
\eeq
where $\omega^{exp}$ is the frequency transfer in the quantum system, measured experimentally,
and $w/J$ is the frequency transfer in units of J from our simulations. 
Note that the critical temperature of the classical Heisenberg model
has been determined from Monte Carlo simulations \cite{kunTc} to be  
$T_c=1.442929(77)J$. Using the normalization for the interaction strength $J$ 
given in Eq. (\ref{jjexp}) and the experimental value $J^{exp}=(6.8\pm 0.6)$K \cite{windsor}
we get $T_c=(85.9\pm 7.6)$K which is well within the error bars of the 
experimental value of around $83$K.

Neutron scattering experiments measure the dynamic structure factor
multiplied by a temperature and frequency dependent population factor 
\cite{coldea,collins,lovesey}, and this factor 
was removed from the experimental data
before comparing them with the simulational data.  Another complication
in the experiment is the finite divergence of the neutron beam which
gives rise to a resolution function which is usually approximated by a
Gaussian in the $4$-dimensional energy and wave-vector space. In the
experiment \cite{coldea}, the measured resolution width along the energy axis 
was $0.25$ meV (full-width at half-maximum) for incoherent elastic scattering.
In order to directly compare our results with the experiment, we convoluted 
the lineshapes from our simulation with a Gaussian resolution function in 
energy with the experimental value of full-width at half-maximum, normalized 
according to Eq. (\ref{wwexp}). The standard deviation $\delta_{\omega}$ 
thus obtained for the Gaussian resolution function in Eq. (\ref{gaussrf}) 
was $0.0619$ in units of $J$.  (Convolution of our lineshapes with the
experimental 3-dimensional Gaussian function in the wave-vector space 
had a negligible effect.)

The experimental data \cite{coldea} are from constant-$q$ scans
with both positive and negative energy transfer. The wave-vector
transfer ${\bf Q}$ was measured along the [1,1,1] direction, around
the antiferromagnetic zone center which in our notation is the $(\pi,\pi,\pi)$
point. Note that Coldea {\it et al}\cite{coldea} define the wave-vector 
transfer ${\bf Q}$ in units such
that the antiferromagnetic zone center is $(0.5,0.5,0.5)$; hence, to express their 
${\bf Q}$ in units of \AA$^{-1}$ one has to multiply it by $2\pi/a$, where $a$ 
is the cubic lattice parameter expressed in \AA. However, in the simulation 
direct lattice positions are defined in units of the lattice constant $a$; 
thus we obtain wave-vectors multiplied by the constant $a$. We emphasize 
that one has to divide the wave-vector ${\bf Q}$ [and also $q$, see 
Eq. (\ref{qpbc})] defined in this paper by $2\pi$ in order to express it in 
the units used by Coldea {\it et al} \cite{coldea}. In the experiment, measurements were 
taken for wave-vectors ${\bf Q}=(\pi+q,\pi+q,\pi+q)$, with $q=2\pi(0.02),$ 
$2\pi(0.04),$..., $2\pi(0.12)$, but unfortunately these values of $q$ are not 
accessible in our simulations for the particular lattice sizes that we 
used. Thus, direct comparison of the lineshapes from the 
simulation with the experimental ones required interpolation of our 
results to obtain the same $q$ values as the experiment. This was done by 
first fitting our lineshapes with a Lorentzian form, as given in 
Eq. (\ref{lorentz}). Since the parameters $B$, $\Gamma_2$ and $\omega_s$ 
obtained from these fits behave as power-laws of $q$, we linearly interpolated 
the logarithm of these parameters as a function of the logarithm of $q$, to 
obtain new parameters for the lineshapes corresponding to those values of 
$q$ actually observed in the experiment.  We estimated the uncertainties from this 
procedure to be less than five percent for the parameter $B$, less than 
three percent for the spin-wave half-width $\Gamma_2$ and the spin-wave 
position $\omega_s$ at $T_c$, and less than one percent for the spin-wave 
position $\omega_s$ at $T=0.9T_c$. Below $T_c$, the parameters $A$ and 
$\Gamma_1$ associated with the central peak were linearly interpolated, 
yielding new values with uncertainties of approximately five percent. 
At $T_c$, the parameter $A$ was interpolated in the log-log plane (as for 
$B$, $\Gamma_2$ and $\omega_s$ discussed above), whereas $\Gamma_1$ was simply linearly 
interpolated. The uncertainties in $A$ and $\Gamma_1$ at $T_c$ were estimated to be less 
than ten percent. For $L=60$, there is one value of $q$, namely $q=2\pi(0.10)$,
which is accessible to both simulation and experiment. This was the only case for which we
did not have to interpolate in $q$. 

Our results at the critical temperature can be compared with the experimental 
data at the same temperature. Below $T_c$, the simulations are mainly for 
$T=0.9T_c$ which unfortunately does not coincide with any temperature used in 
the experiment; however, it is very close to $T=0.894T_c$ 
for which experimental results are available.  To correct for the 
slight difference, we made a linear interpolation in temperature, using our 
results for $L=24$ at $T=0.846T_c$ and at $T=0.9T_c$. We first fitted the lineshapes at 
these two temperatures to the form given by Eq. (\ref{lorentz}), then we
linearly interpolated the position and the amplitude of the spin-wave peak at these 
temperatures, to obtain the spin-wave position and amplitude corresponding to $T=0.894T_c$. 
For small values of $q$ we found that the frequency and amplitude of the spin-wave peak at 
$T=0.894T_c$ were respectively $\sim 1.5$ and $5$ percent larger than at $T=0.9T_c$ and this 
difference decreased for larger values of $q$. 

The intensity of the lineshapes in the neutron scattering experiment was 
measured in counts per 15 seconds. For both temperatures $T=0.894T_c$ and 
$T=T_c$ the measurements for the several wave-vectors were done with the 
same experimental set-up and conditions.  Therefore, the relative 
intensities of the lineshapes for the different wave-vectors is fixed,
and equal for both temperatures. The intensity of the lineshapes obtained in the simulation 
had to be normalized to the experimental value; however, because the relative intensities for 
different wave-vectors is fixed, we have only one independent normalization factor for all the 
wave-vectors at both temperatures. The normalization of the intensity was chosen so that
the spin-wave peak for $T=0.894T_c$ and $q=2\pi(0.08)$ from the experiment and the simulation 
matched. This same factor was used to normalize the intensities of the lineshapes corresponding 
to the remaining values of wave-vectors at $T=0.894T_c$, and for all cases 
at $T_c$. 

\begin{figure}
\vspace{-2.0cm}
\centering
\leavevmode
\epsfxsize=2.8in
\begin{center}
\leavevmode
\epsffile{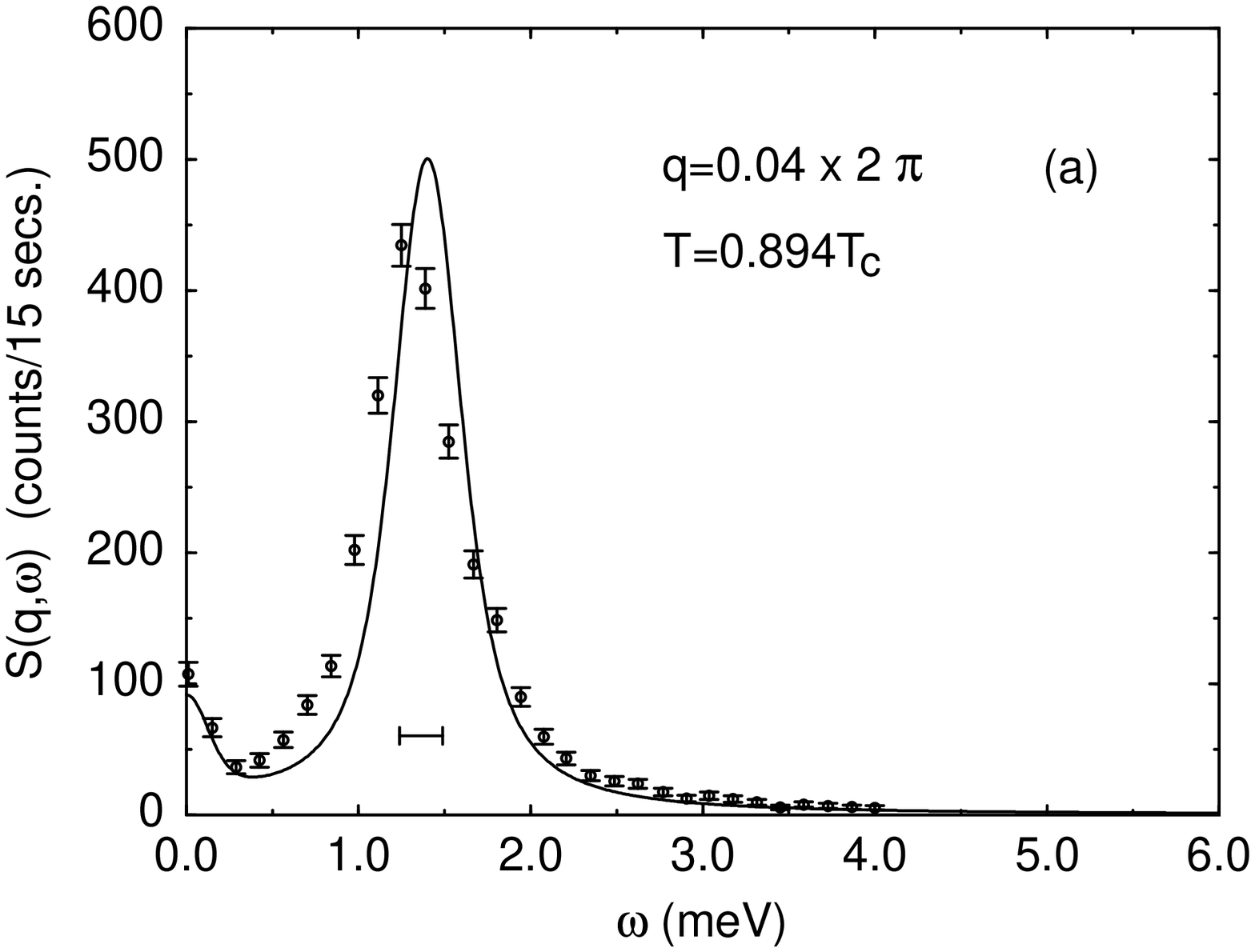}
\end{center}
\vspace{-3.3cm}
\begin{center}
\leavevmode
\epsfxsize=2.8in
\epsffile{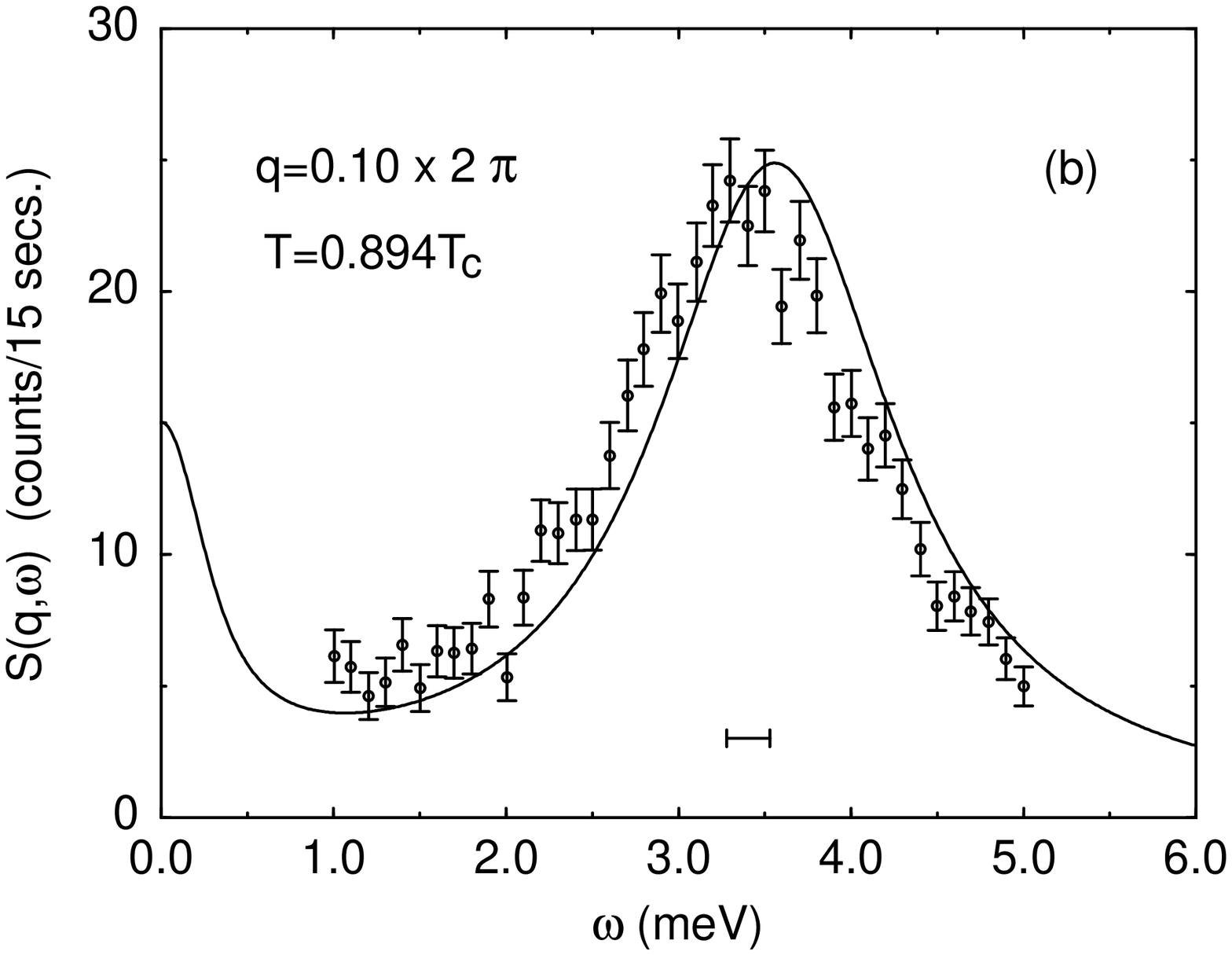}
\end{center}
\vspace{-1.3cm}
\caption{Comparison of lineshapes obtained from fits to
simulation data for $L=60$ (solid line) and experiment (dots) at $T=0.894T_c$ in 
the [111] direction. The horizontal line segment in each graph 
shows the resolution in energy (full-width at half-maximum).}
\label{ceTp894}
\end{figure}

\begin{figure}
\vspace{-2.2cm}
\centering
\leavevmode
\epsfxsize=3.0in
\begin{center}
\leavevmode
\epsffile{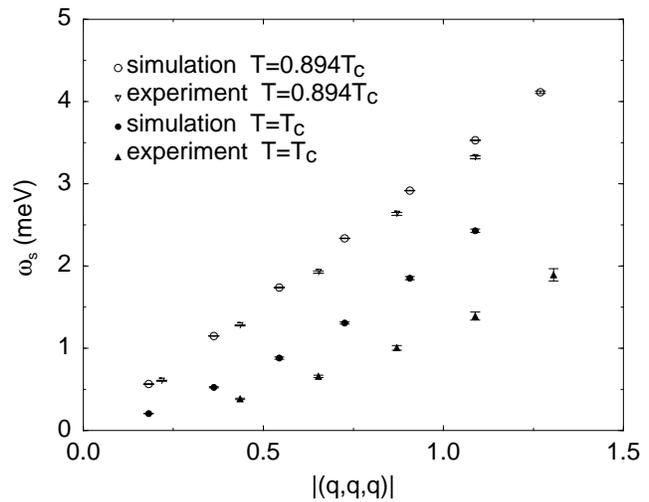}
\end{center}
\vspace{-1.3cm}
\caption{Comparison of the dispersion curve at $T=T_c$ and $T=0.894T_c$ 
obtained from simulations for $L=60$ 
(circle) and the experiment (triangle), in the [111] direction. The simulation
data shown here correspond to values of $q$ accessible with $L=60$, without interpolation
to match the $q$ values from the experiment. In this notation, the first Brillouin zone 
edge is at $|(q,q,q)|\simeq 2.72$.}
\label{cewqTp9}
\end{figure}

\begin{figure}
\vspace{-3.0cm}
\centering
\leavevmode
\epsfxsize=3.2in
\begin{center}
\leavevmode
\epsffile{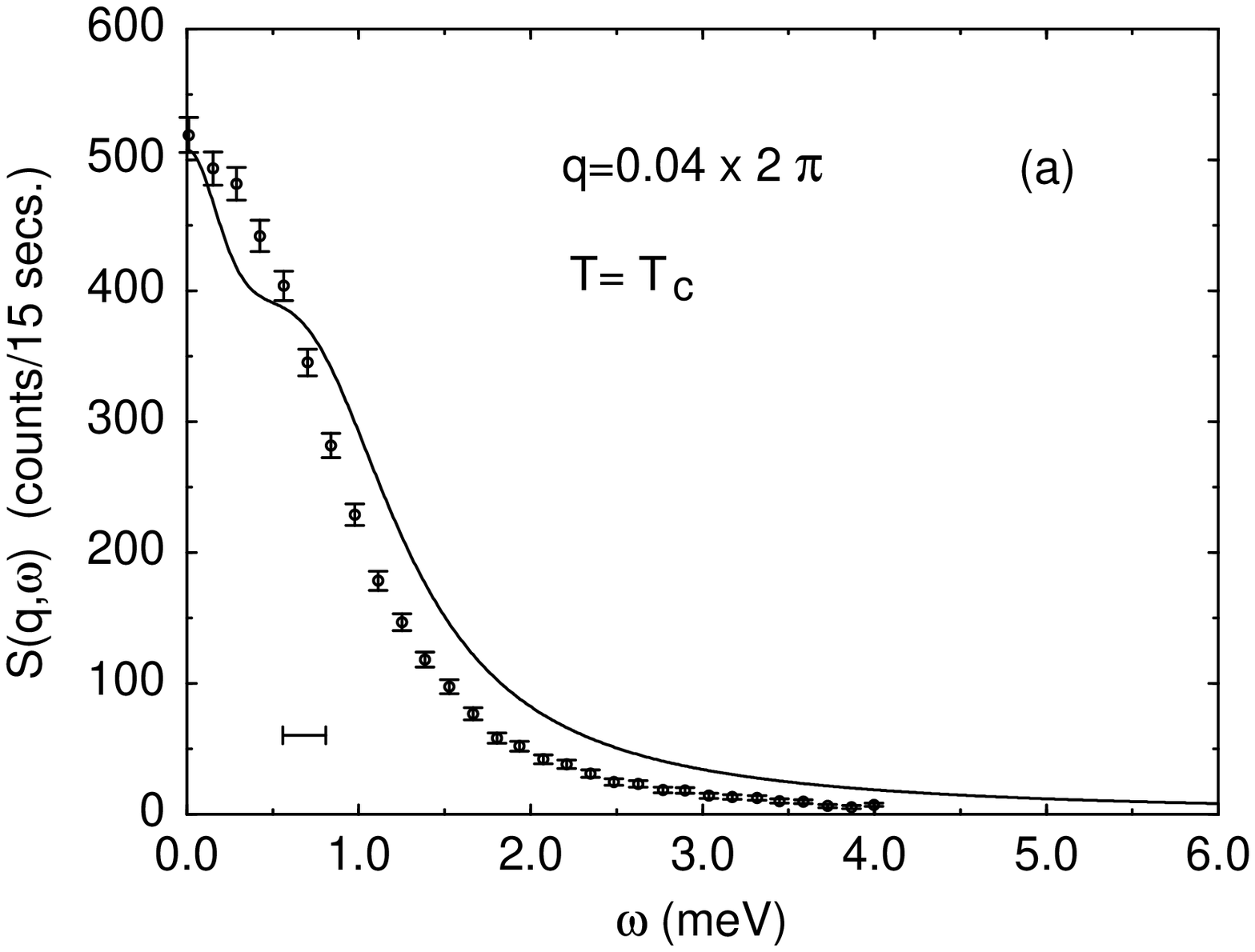}
\end{center}
\vspace{-4.5cm}
\begin{center}
\leavevmode
\epsfxsize=3.2in
\epsffile{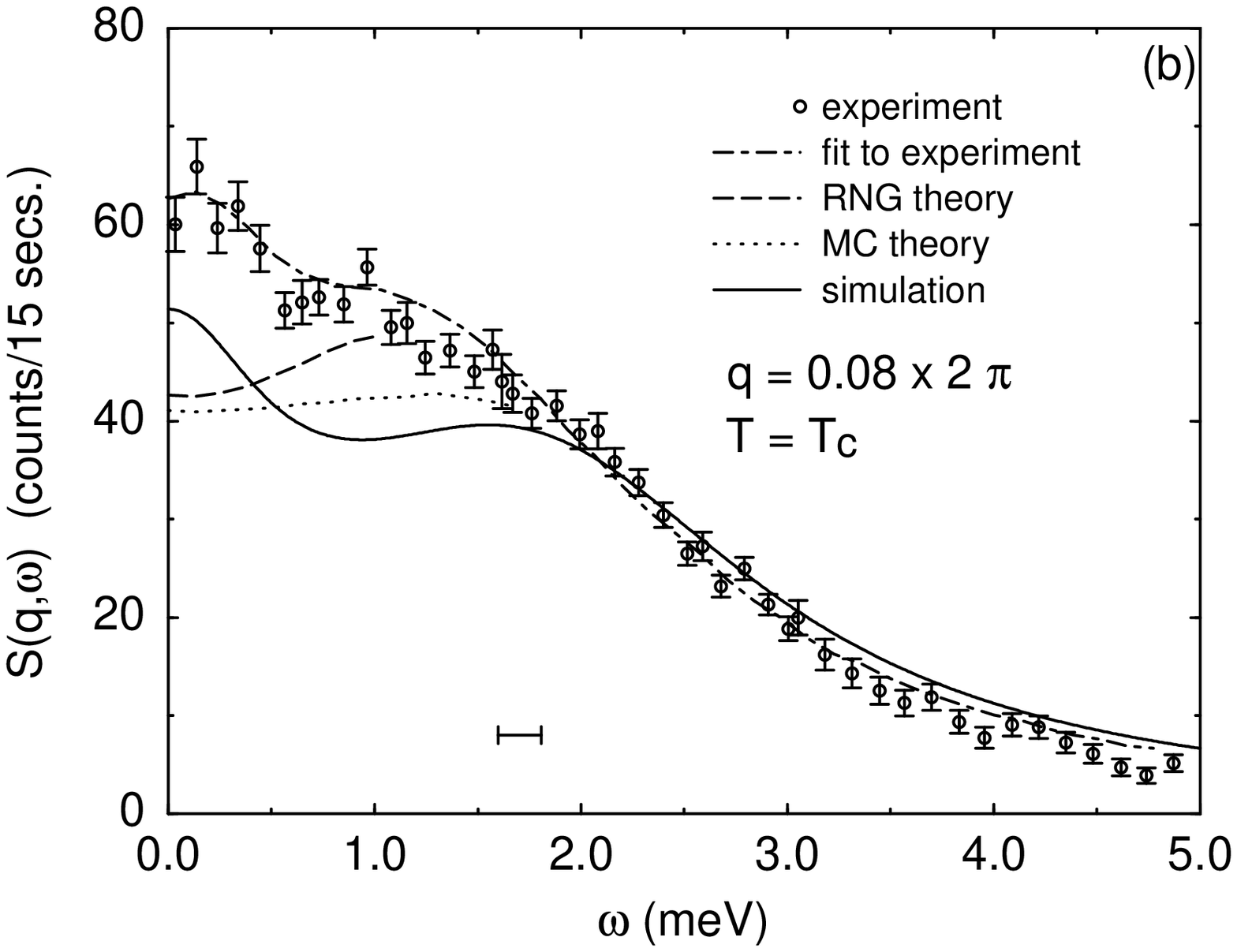}
\end{center}
\vspace{-1.1cm}
\caption{Comparisons of lineshapes obtained from fits to
simulation data for $L=60$ (solid line) and the experiment (open circles) at $T=T_c$ in the [111] 
direction, for (a) $q=2\pi(0.04)$ and (b) $q=2\pi(0.08)$. 
The dot-dashed line in (b) is a fit of the experimental data to the 
functional form predicted by the RNG theory plus a Lorentzian central peak, and the RNG 
component of the fit is shown by the long-dashed line. The prediction of Mode Coupling (MC) theory 
is shown by the dotted line in (b). The horizontal line segment in each graph 
represents the resolution in energy (full-width at half-maximum).}
\label{ceTc}
\end{figure}

The final lineshapes for $T=0.894T_c$, $L=60$, and two wave-vectors are 
shown in Fig. \ref{ceTp894} together with experimental lineshapes for each case.
Fig. \ref{cewqTp9} shows the comparison of the 
dispersion curve from the simulation and the experiment at 
$T=0.894T_c$. Good agreement between our results and experiment can be
seen from either the direct comparison of the lineshapes, or the comparison of the dispersion
curve. The lineshape intensities from simulation and experiment agree over two orders of 
magnitude, from $q=2\pi(0.02)$ to $q=2\pi(0.10)$.
Fig. \ref{ceTc} shows the comparison of lineshapes from the simulation and the experiment for
$T=T_c$, $L=60$, and 
%
%\vspace{1.5cm}
%
%\vspace{2.0cm}
$\;$ two values of $q$. The dispersion curve obtained from the simulations 
at $T=T_c$, shown in Fig. \ref{cewqTp9}, is systematically larger than the experimental 
values.  We emphasize that the error bars shown for the dispersion curve
obtained from our simulations at $T_c$ reflect only the statistical
errors of a best fit of the lineshapes. For each $q$, this fit was done 
with only one range of frequency; hence errors associated with the choice 
of frequency range and the quality of the fit were not taken into account. It is 
reasonable to expect that such sources of error would increase the error bars by a factor of 5.
 From the direct comparison of the simulated and experimental lineshapes at $T_c$ it is difficult 
to determine the difference in the spin-wave frequencies, because the 
spin-wave peaks from the experiment are not very pronounced, and their 
positions have to be extracted from the fits of 
the lineshapes. As mentioned earlier, the experimental data at $T_c$ were
fitted to a functional form predicted by RNG theory plus a Lorentzian
central peak. The quality of the fitting can be seen in Fig. \ref{ceTc}(b)
for $q=2\pi(0.08)$, along with the RNG component of the fit and the
prediction by mode-coupling theory. Finally, even though at $T_c$ the
lineshape intensities from the simulations for small frequency
transfer tended to be lower as compared to the experiment, the
agreement is still reasonably good, considering the variation of the
intensities over almost two orders of magnitude from $q=2\pi(0.02)$
to $q=2\pi(0.12)$. 

\section{Conclusion}

We have studied the magnetic excitations and the dynamic critical
properties of the classical Heisenberg antiferromagnet
on a simple cubic lattice using large-scale spin dynamics simulations.
A new 4th-order decomposition integration technique as implemented
in Ref.\citen{krech}  allowed us to use a larger time integration step
thus extend the maximum integration time to much larger
values than was previously possible.

Below $T_c$, the dispersion curves were approximately linear for small $q$. 
Increasing the temperature towards the critical
temperature the dispersion curve became a power-law, reflecting the crossover from
hydrodynamic to critical behavior. The spin-wave half-width at
$T=0.9T_c$ also showed a crossover from critical behavior at large values of $q$ to 
hydrodynamic behavior at small values of $q$. The dynamic critical exponent was estimated
to be $z=(1.43\pm 0.03)$ which is in agreement with the experimental value of Coldea {\it et al}
\cite{coldea} and slightly lower than the dynamic scaling prediction. 

We made direct, quantitative comparison of both the dispersion curve and the
lineshapes obtained from our simulations with the recent experimental results
by Coldea {\it et al}\cite{coldea}  for RbMnF$_3$. At $T=0.894T_c$ the agreement was
quite good with the major difference being at $T_c$ where spin-wave peaks
from our simulations tended to be at slightly larger frequencies than
the experimental results. Both at $T=0.894T_c$ and at $T_c$ the
lineshape intensities varied over almost two orders of magnitude from
$q=2\pi(0.02)$ to $q=2\pi(0.10)$ and there was good agreement
between the intensities from simulation and experiment over the entire range.
Thus, the simple isotropic nearest-neighbor classical Heisenberg model
describes the dynamic behavior of this real magnetic system quite well.

\begin{center}
{\bf ACKNOWLEDGMENTS}
\end{center}

We thank R. A. Cowley and R. Coldea for helpful discussions
and for sending us their data. We also thank M. Krech and  
H. B. Sch\"{u}ttler for valuable discussions. Simulations were carried out on 
the Cray T90 at the San Diego Supercomputing Center, and on a SGI Origin2000 
and IBM R6000 machines in the University of Georgia. This research was supported in part 
by NSF Grant No. DMR-9727714.

\end{document}